\documentclass[letterpaper,english,reprint,aps]{revtex4-1}
\usepackage[T1]{fontenc}
\usepackage[latin9]{inputenc}
\usepackage{color}
\usepackage{textcomp}
\usepackage{amsmath}
\usepackage{amssymb}
\usepackage{graphicx}
\usepackage[normalem]{ulem}

\makeatletter

\pdfpageheight\paperheight
\pdfpagewidth\paperwidth

\usepackage{babel}

\makeatother

\usepackage{babel}
\begin{document}
\title{Analytical Quantitative Semi-Classical Approach to the LoSurdo-Stark Effect and Ionization
in 2D excitons}
\author{J. C. G. Henriques$^{1}$, H{\o}gni C. Kamban$^{2,3}$, Thomas G. Pedersen$^{2,3}$,
N. M. R. Peres$^{1,4}$}
\affiliation{$^{1}$Department and Centre of Physics, and QuantaLab, University
of Minho, Campus of Gualtar, 4710-057, Braga, Portugal}
\affiliation{$^{2}$Department of Materials and Production, Aalborg University, DK-9220 Aalborg \O st, Denmark }
\affiliation{$^{3}$ Center for Nanostructured Graphene (CNG), DK-9220 Aalborg \O st, Denmark}
\affiliation{$^{4}$International Iberian Nanotechnology Laboratory (INL), Av. Mestre
Jos{\'e} Veiga, 4715-330, Braga, Portugal}
\date{\today}
\begin{abstract}
Using a semi-classical approach, we derive a fully analytical expression
for the ionization rate of excitons in two-dimensional materials due
to an external static electric field, which eliminates the need for complicated numerical calculations. Our formula shows quantitative
agreement with more sophisticated numerical methods based on the exterior complex scaling approach, which solves a non-hermitian eigenvalue problem yielding complex energy eigenvalues, where the imaginary part describes the ionization rate.
Results for excitons in hexagonal boron nitride and the $A-$exciton in transition metal dichalcogenides are given as a simple examples. The extension of the theory to include spin-orbit-split excitons
in transition metal dichalcogenides is trivial. 
\end{abstract}
\maketitle

\section{Introduction\label{sec:Introduction}}

The LoSurdo-Stark effect has a venerable history in both atomic, molecular,
and condensed matter physics \citep{connerade}. This effect refers
to the modification of the position of the energy levels of a quantum
system due to the application of an external electric field and, in
addition, to the possible ionization of atoms, molecules,
and excitons due to the very same field. The latter effect is a
nice example of quantum tunneling through an electrostatic barrier.
The difference between values of the energy levels with and without
the field is dubbed the Stark shift. The ionization process is characterized
by an ionization rate, which depends on the magnitude of the external electric field, as well as material parameters. Although the calculation
of the Stark shift can be easily accomplished using perturbation theory
\citep{Franceschini1985}, the calculation of the ionization rate
is non-perturbative \citep{nicolaides1992} since it is proportional
to $\exp(-\beta/F)$, where $\beta$ is a material-dependent parameter
and $F$ is the magnitude of the external electric field. For the
case of the Hydrogen atom in three dimensions the literature on the
LoSurdo-Stark effect, spanning a period of about 100 years, is vast.
On the contrary, for low dimensional systems, such as the two-dimensional
Hydrogen atom, the calculation of both the Stark shift and the ionization
rate for very strong electric fields was only recently considered
 \citep{Pedersen2016, Mera2015}; the weak field
limit had been studied by Tanaka \emph{et al} prior \citep{Tanaka1987}. 
In the previous work, the authors used a low order perturbation expansion of the energy, combined with the hypergeometric resummation technique \cite{Mera2015,Pedersen2016} to extract the full non-perturbative behavior of the enregy and thus address the LoSurdo-Stark
effect in a system they dubbed low-dimensional Hydrogen. Other numerical methods for tackling the
calculation of Stark shifts and the ionization rates include the popular complex scaling method \cite{Herbst1978}, as well as the
Riccati-Pad{\'e} method \citep{Fernandez2018}, which is based on the transformation
of the Schr\"{o}dinger equation to a nonlinear Riccati equation. Also,
using the same mapping, Dolgov and Turbiner devised a numerical perturbative
method \citep{dolgov1980}, starting from an interpolated solution
of the Riccati equation, for computing the ionization rate at strong
fields. The LoSurdo-Stark problem has also been addressed using JWKB
schemes \citep{Rice1962,Bekenstein1969,gallas1982,harmin1982} and
variational methods \citep{Froelich1975}. Fully analytical results
have been found for the three dimensional Hydrogen atom \citep{Yamabe1977}.
Another interesting approach uses Gordon-Volkov wave functions \citep{Joachain2012,Schultz2014},
which are semi-classical-type wave functions for an electron in
the electric field of an incoming electromagnetic wave.

In the field of two-dimensional (2D) excitons \citep{Gang2018ColloquiumExcitons},
there are already experimental reports of both valley selective Stark
shift \citep{Sie:2015aa,Sie2016,Cunningham:2019aa} and exciton dissociation
via external electric fields \citep{Massicotte:2018aa}. In the same
context, Stark shifts and ionization rates of excitons in these condensed matter systems have been calculated theoretically for arbitrary field intensities \citep{pedersen2016b,Scharf:2016,Kamban2019}. In \cite{Kamban2019} a semi-analytical method was used, where the field-dependence of these
two quantities (shift and ionization rate) are determined
analytically, and a material-dependent constant is determined numerically.
The method only requires the electrostatic potential to have a Coulomb $1/r$ 
tail, but involves the introduction of an extra basis of functions
to deal with the non-separability of the electrostatic potential.
Their results are asymptotically exact and capture commendably the
low-field regime. The same authors have recently extended the method
to excitons in van der Waals heterostructures \citep{das2015beyond} with
success \citep{Kamban2020}. Recently, Cavalcante \emph{et al. }have
extended the interest in the Stark effect to trions in 2D semiconductors
\citep{Cavalcante2018}.

At the time of writing, there is not a fully analytical description
of the LoSurdo-Stark effect for excitons in 2D materials. Although
the previous methods can be used to describe this effect, the lack
of a fully analytical expression prevents their use by a wider community
and lacks the insight provided by an analytical description. This
is especially true for the material-dependent coefficient which is both
hard to obtain numerically and varies by orders of magnitude even for modest
changes of the dielectric function of the materials encapsulating
the 2D material. An additional difficulty is the non-separability
of the electrostatic, non-Coulombic, potential between the hole and
the electron in 2D excitons. This non-separability has hindered the
use of well known methods based on parabolic coordinates \citep{Alexander1969,Castillo2008,Damburg1978,dolgov1980}.
As we will see, however, this difficulty may be circumvented if one introduces
the concept of an effective potential \citep{Pfeiffer2012}, with
the only requirement being the existence of a Coulomb tail at large
distances in the electrostatic potential \citep{Bisgaard2004,lars2011}.
This concept, in essence, renders the original potential approximately
separable if one focuses on the relevant coordinate. Indeed, in parabolic
coordinates, $\xi$ and $\eta$, and for the Hydrogen atom in 2D in
a static electric field the eigenvalue problem is separable. The two
resulting equations describe two different types of quantum problems.
Whereas in the $\xi$ coordinate the eigenvalue problem is that of
a bound state, in the $\eta$ coordinate the resulting eigenvalue
problem describes a scattering state, where the exciton dissociates
via tunneling through the Coulomb barrier, the latter rendered
finite by the presence of the static electric field. Since tunneling
is the relevant mechanism for dissociation and occurs (for weak
fields) at large values of $\eta$, the problem, which is initially
non-separable, effectively becomes a function of one of the coordinates
alone, depending on which of the two eigenvalue problems we are considering.

In this paper we take advantage of a number of techniques and obtain
a fully analytical formula for the non-perturbative ionization rate
of 2D excitons. Our approach highlights the role of both the excitons'
effective mass and the dielectric environment, providing a simple
formula for the ionization rate, in full agreement with more demanding numerical
methods for weak fields. Such a formula is very useful for quick estimates of the ionization rate of excitons in 2D materials, and provides physical intuition that is helpful in, e.g., device design.

This paper is organized as follows: in Sec. \ref{sec:Schrodinger-Equation}
we present the Wannier equation describing the relative motion of
an electron-hole pair and discuss the approximate separability of the Rytova-Keldysh
potential. In Sec. \ref{sec:Ionization-Rate} we obtain the expression
for the ionization rate and discuss the semi-classical solution of
the tunneling problem. In Sec. \ref{sec:Results} we apply our formula
to the calculation of the ionization rate of excitons in 2D materials,
taking the examples of hexagonal boron nitride and transition metal dichalcogenides. Finally, in Sec. \ref{sec:Conclusions}
we give some final notes about our work and possible extension of
the results.

\section{Wannier Equation\label{sec:Schrodinger-Equation}}

\begin{figure}
\includegraphics[width=8cm]{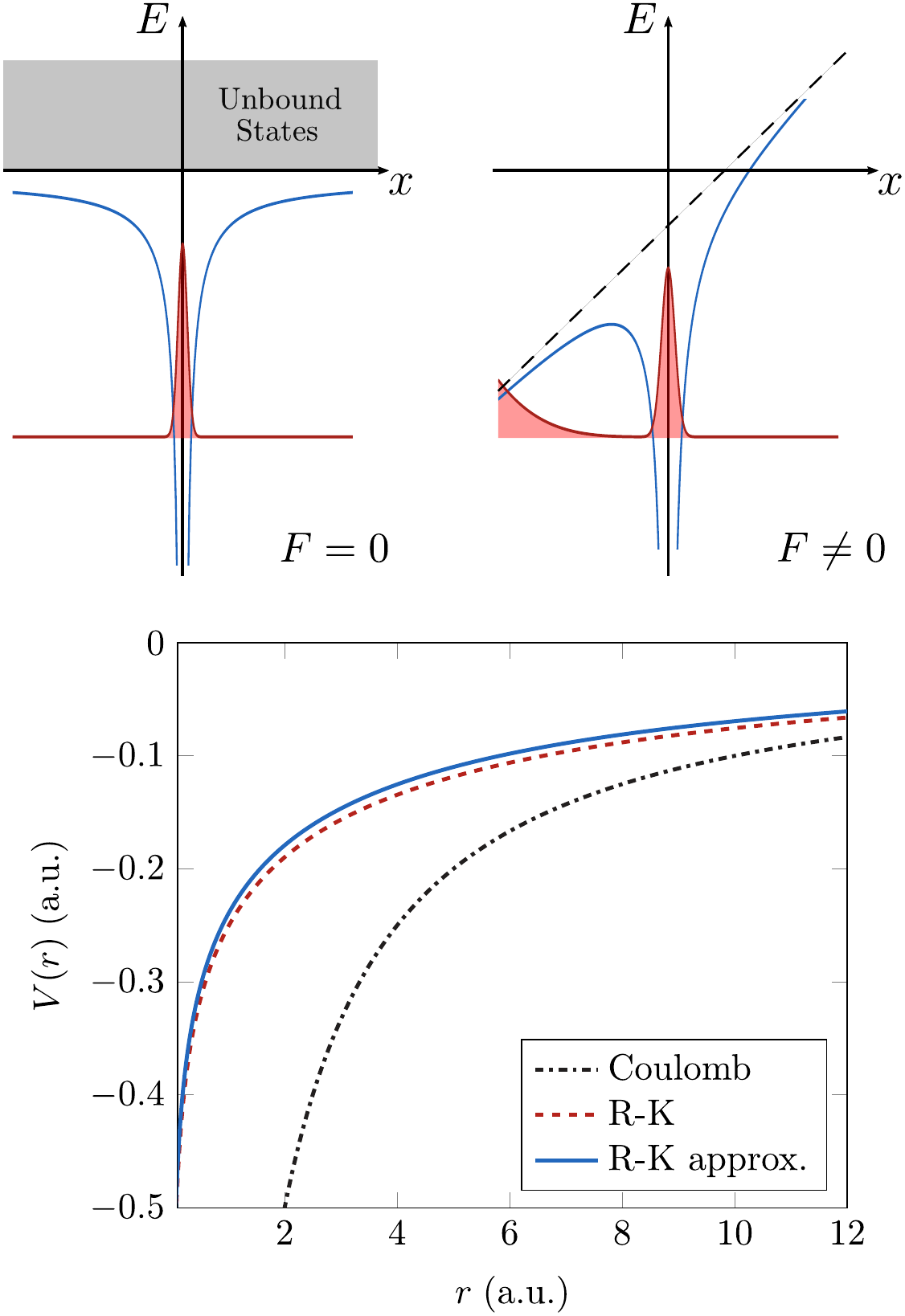}
\caption{(Top) Electrostatic potential and wave functions of the LoSurdo-Stark effect.
In the left panel the system is represented in the absence of the
electric field and a bound state is formed. In the right panel an
external electric field is superimposed on the attractive electrostatic
field distorting the latter. Along one of the directions the total
potential becomes more confining whereas in the opposite direction
the external field creates  a barrier through which the exciton
can tunnel and thereby ionize. (Bottom) Comparison between the Coulomb potential and the two expressions defined in the text for the Rytova-Keldysh potential. While for large values of $r$ the potentials present an identical behaviour, for small $r$ a significant difference between the Coulomb and Rytova-Keldysh potentials is visible. Moreover, it is clear that the approximate expression for the Rytova-Keldysh potential gives an excellent approximation of Eq. (\ref{eq:RK potential full}). The parameters $\kappa=1$ and $r_0 = 10$ {\AA} were used. \label{fig:Electrostatic-potential-and}}
\end{figure}

In this section we introduce the Wannier equation, originating from
a Fourier transform of the Bethe-Salpeter equation \citep{Henriques2020hBN,Henriques2020Phosphorene,Henriques2020TMD},
that defines the exciton problem in real space. We have found in previous publications
\citep{Have2019,Henriques2020hBN,Henriques2020Phosphorene,Henriques2020TMD} a
good agreement between the solution of the Bethe-Salpeter equation
and the binding energies arising from the solution of the Wannier
equation. The physics of the LoSurdo-Stark effect is qualitatively
represented in Fig. \ref{fig:Electrostatic-potential-and}.

Following the steps of \citep{Alexander1969,Damburg1978,Castillo2008}
we will pass from polar to parabolic coordinates with the goal of decoupling
the original two-variable differential equation into two one-dimensional differential equations. Unlike the pure Coulomb problem, this problem is not exactly separable. However, as will become apparent, this problem is separable under justifiable approximations. 

In this work we are interested in studying the ionization rate of
excitons in 2D materials due to an external static electric field.
The Wannier equation in atomic units (a. u.) and in terms of the relative coordinate reads 
\begin{equation}
\nabla^{2}\psi(\mathbf{r})-2\mu\left[-E+\mathbf{F\cdot r}+V(\mathbf{r})\right]\psi(\mathbf{r})=0,\label{eq:Wannier Eq.}
\end{equation}
where $\mu$ is the reduced mass of the electron-hole system, $E$
is the energy, and $\mathbf{F}=F \hat{x}$, with $F>0$, the external electric
field, that we consider aligned along the $x$ direction. The electron-hole interaction $V(\mathbf{r})$ is given by the Rytova-Keldysh potential \citep{rytova1967,keldysh1979coulomb}
\begin{equation}
V(\mathbf{r})=-\frac{\pi}{2r_0}\left[\mathbf{H}_{0}\left(\frac{\kappa r}{r_{0}}\right)-Y_{0}\left(\frac{\kappa r}{r_{0}}\right)\right],\label{eq:RK potential full}
\end{equation}
where the so-called screening length $r_0$ is proportional to the polarizability of the 2D sheet \cite{Cudazzo2011}. Macroscopically, it may be related to the thickness $d$ and dielectric function $\epsilon$ of the sheet as $r_{0}\sim d\epsilon/2$. Furthermore, $\kappa$ is the mean dielectric constant of the media
above and below the 2D material, $\mathbf{H}_{0}$ is the Struve
function, and $Y_{0}$ is the Bessel function of the second kind. The fact that the electrostatic interaction between electron-hole pairs in 2D materials is given by the Rytova-Keldysh potential is what gives rise to the nonHydrogenic Rydberg series \citep{Chernikov2014}. 
Inspired by the work of \citep{Cudazzo2011}, where an approximate
expression for the Rytova-Keldysh potential is presented, we use the
following expression as an approximation to Eq. (\ref{eq:RK potential full})
\begin{equation}
V(\mathbf{r})\approx\frac{1}{r_{0}}\log\frac{\kappa r}{\kappa r+r_{0}}.
\end{equation}
In Fig. \ref{fig:Electrostatic-potential-and} we plot Eq. (\ref{eq:RK potential full}) and the previous expression
in the same graph and observe that the latter formula is an excellent  approximation of the former.

Next we note that several authors
\citep{Alexander1969,Damburg1978,Castillo2008} used parabolic coordinates
in order to separate the Schr\"{o}dinger equation into two differential
equations of a single variable. In those works, however, the Coulomb
potential was considered in the 3D case. In our case, the Rytova-Keldysh
potential does not allow a simple solution by separation of variables.
To be able to do so, an effective potential has to be introduced.
Let us consider the following set of parabolic coordinates \citep{Joana2016,Hogni2018MSc}
in 2D: 
\begin{align}
x & =\frac{\xi-\eta}{2},\label{eq:x_coord}\\
y & =\pm\sqrt{\xi\eta},\label{eq:y_coord}\\
r & =\frac{\xi+\eta}{2},\label{eq:r_coord}
\end{align}
with both $\xi$ and $\eta$ belonging to the interval $[0,\infty[$. In these new coordinates the Laplacian reads
\begin{equation}
\nabla^{2}=\frac{4}{\eta+\xi}\left[\sqrt{\eta}\frac{\partial}{\partial\eta}\left(\sqrt{\eta}\frac{\partial}{\partial\eta}\right)+\sqrt{\xi}\frac{\partial}{\partial\xi}\left(\sqrt{\xi}\frac{\partial}{\partial\xi}\right)\right].
\end{equation}
Applying this variable change to Eq. (\ref{eq:Wannier Eq.}), and
considering that $\psi(\eta,\xi)=v(\eta)u(\xi),$ we obtain an equation
that can be separated, except for the potential term where $\xi$
and $\eta$ are still coupled by
\begin{equation}
\frac{(\eta+\xi)}{2}V\left(\frac{\eta+\xi}{2}\right).
\end{equation}
To fully separate the $\xi$ and $\eta$ dependencies we propose the
following effective potential 
\begin{equation}
\frac{(\eta+\xi)}{2}V\left(\frac{\eta+\xi}{2}\right)\approx \frac{\eta}{2} V\left(\frac{\eta}{2}\right)+\frac{\xi}{2} V\left(\frac{\xi}{2}\right).
\end{equation}
The reasoning behind this choice is as follows: we know that in the
usual polar coordinates the Rytova-Keldysh potential obeys the following
two limits
\begin{align}
\lim_{r\rightarrow0}rV(r) & =0,\\
\lim_{r\rightarrow\infty}rV(r) &= -\frac{1}{\kappa}.
\end{align}
It is therefore clear that the decoupling we have introduced respects
the two previous limits. We have chosen the above separation for having
asymptotically the Coulomb potential in both $\eta$ and $\xi$ coordinates.
The quality of the approximation has to be judged by the accuracy of the formula
for the ionization rate (anticipating the results, we find an excellent
qualitative and quantitative agreement between the analytical results
and the numerical ones). In view of the approximation made, the decoupled
equations read
\begin{align}
\left[\sqrt{\eta}\frac{\partial}{\partial\eta}\left(\sqrt{\eta}\frac{\partial}{\partial\eta}\right)+\frac{\mu E}{2}\eta-\frac{\mu}{2}\eta V\left(\frac{\eta}{2}\right)+\mu\frac{F}{4}\eta^{2}-Z\right]u(\eta) & =0,\\
\left[\sqrt{\xi}\frac{\partial}{\partial\xi}\left(\sqrt{\xi}\frac{\partial}{\partial\xi}\right)+\frac{\mu E}{2}\xi-\frac{\mu}{2}\xi V\left(\frac{\xi}{2}\right)-\mu\frac{F}{4}\xi^{2}+Z\right]v(\xi) & =0,
\end{align}
where $Z$ was introduced as a separation constant. Its value is determined
below, demanding the correct large distance asymptotic behavior of
the wave function. Using the proposed effective potential, the equations
become mathematically identical, as they should, except for the external field
term. Moreover, both equations reproduce the Coulomb problem in the
asymptotic limit. We note that the two quantum problems defined in
terms of the $\xi$ and $\eta$ coordinates have a completely different
nature. While the $\xi-$equation defines a bound state problem, and is therefore tractable by simple methods, the $\eta-$equation defines a
tunneling problem and obtaining its exact solution is challenging. These differential
equations can be further simplified with the introduction of
\begin{align}
u(\eta) & =\frac{u_{1}(\eta)}{\eta^{1/4}},\\
v(\xi) & =\frac{v_{1}(\xi)}{\xi^{1/4}},
\end{align}
which leads to
\begin{align}
\left(\frac{d^{2}}{d\eta^{2}}+\frac{3}{16\eta^{2}}+\frac{\mu E}{2}-\frac{\mu}{2}V\left(\frac{\eta}{2}\right)+\frac{\mu F}{4}\eta-\frac{Z}{\eta}\right)u_{1}(\eta) & =0\label{eq:v1 eta diff eq}\\
\left(\frac{d^{2}}{d\xi^{2}}+\frac{3}{16\xi^{2}}+\frac{\mu E}{2}-\frac{\mu}{2}V\left(\frac{\xi}{2}\right)-\frac{\mu F}{4}\xi+\frac{Z}{\xi}\right)v_{1}(\xi) & =0.\label{eq:u1 xi diff eq}
\end{align}
As noted, solutions of these two equations at large distances from
the origin have two different behaviors: the first one has an oscillatory
behavior, whereas the second one decays exponentially. Furthermore,
while the second equation has a discrete spectrum, the first has a
continuous one. This is due to the different sign of the field term in the two
equations.

Finally, to end this section, we introduce another change of variable
that has already proven to be of great value in this type of problems,
known as the Langer transformation \citep{langer1937connection,Good1953,berry1973semiclassical,Farrelly_1983},
which is defined by
\begin{align}
\eta & =e^{t}\Rightarrow d\eta=e^{t}dt,\\
u_{1}(\eta) & =e^{t/2}T(t).\label{eq:Langer v1 T}
\end{align}
Making use of this transformation, Eq. (\ref{eq:v1 eta diff eq})
acquires the form 
\begin{equation}
T''(t)+P^{2}(t)T(t)=0,\label{eq:T(t) diff eq}
\end{equation}
with 
\begin{equation}
P^{2}(t)=\frac{\mu E}{2}e^{2t}+\frac{\mu F}{4}e^{3t}-Ze^{t}-\frac{\mu}{2}V\left(\frac{e^{t}}{2}\right)e^{2t}-\frac{1}{16}\label{eq:P squared def}
\end{equation}
A similar transformation could be applied to Eq. (\ref{eq:u1 xi diff eq}).
This, however, is not necessary. We note that the advantages of the
above transformation are two-fold: on one side, the initial problem,
valid only for $\eta\in[0,\infty[$, is transformed into a one-dimensional
problem in the interval $t\in]-\infty,\infty[$, and on the other
it removed the singular behavior at the origin due to the terms associated
with $3/16\xi^2.$ As shown by Berry and Ozorio de Almeida \citep{berry1973semiclassical},
the Langer transformation is a key step in solving the 2D Hydrogen
problem for zero angular momentum, which is similar to the problem
at hands.

\section{Ionization Rate\label{sec:Ionization-Rate}}

In this section we will present a derivation for the ionization rate
of excitons in 2D materials due to the external electric field. We
will start by associating the ionization rate with an integral of
the probability current density. This integral will contain the function
$T(t)$, presented in the end of the previous section. Then, this
function will be explicitly computed. Afterwards, combining the previous
two steps, a fully analytical expression for the ionization rate will
be presented.

\subsection{The ionization rate formula}

In the beginning of the text we considered the electric field to be
applied along the $x$ direction, implying that the electrons will
escape via the negative $x$ direction, which in the parabolic coordinates
introduced in Eqs. (\ref{eq:x_coord})-(\ref{eq:r_coord}) corresponds
to large values of $\eta.$ Since the electrons will escape along
the negative $x$ direction, we can define the ionization rate $W$
as \citep{Bisgaard2004}
\begin{equation}
W=-\int_{-\infty}^\infty j_{x}dy.\label{eq:Ionization rate}
\end{equation}
That is, the number of particles per unit time transversing a line
perpendicular to the probability current density $j_{x},$ which reads
\begin{equation}
j_{x}=\frac{i}{2\mu}\left(\psi\frac{\partial\psi^{*}}{\partial x}-\psi^{*}\frac{\partial\psi}{\partial x}\right),\label{eq:current density jx}
\end{equation}
where, once again, $\mu$ is the reduced mass. In terms of parabolic
coordinates, the position vector is
\begin{equation}
\mathbf{r}=\frac{\xi-\eta}{2}\hat{\mathbf{x}}+\sqrt{\xi\eta}\hat{\mathbf{y}}.
\end{equation}
The differentials in Cartesian coordinates are related to the parabolic
ones through the following relations
\begin{align}
dx & =\frac{1}{2}d\xi-\frac{1}{2}d\eta,\\
dy & =\frac{1}{2}\sqrt{\frac{\eta}{\xi}}d\xi+\frac{1}{2}\sqrt{\frac{\xi}{\eta}}d\eta.
\end{align}
From here we find
\begin{align}
\frac{\partial}{\partial x} =\frac{\partial\xi}{\partial x}\frac{\partial}{\partial\xi}+\frac{\partial\eta}{\partial x}\frac{\partial}{\partial\eta}
 \approx-2\frac{\partial}{\partial\eta},
\end{align}
where the final approximation comes from considering the limit $x \rightarrow -\infty$. Recalling what was done in the previous
section we write
\begin{align}
\psi(\eta,\xi) & = u(\eta) v(\xi)\\
 &=\frac{u_{1}(\eta)}{\eta^{1/4}}\frac{v_{1}(\xi)}{\xi^{1/4}}.
\end{align}
Employing Eq. (\ref{eq:Langer v1 T}) this may also be written as
\begin{equation}
\psi(\eta,\xi)=\eta^{1/4}T\left[t(\eta)\right]\frac{v_{1}(\xi)}{\xi^{1/4}}.
\end{equation}
The probability current density introduced in Eq. (\ref{eq:current density jx})
can now be computed in parabolic coordinates as
\begin{equation}
j_{x}\approx\frac{i}{\mu}\sqrt{\eta}\frac{\vert v_{1}(\xi)\vert^{2}}{\xi^{1/2}} \left(T^{*}\frac{dT}{d\eta}-T\frac{dT^{*}}{d\eta}\right),
\end{equation}
where, following the same reasoning as before, the derivatives in
$\xi$ were ignored. Inserting this expression into Eq. (\ref{eq:Ionization rate})
and approximating the differential in the $y$ coordinate by $dy\approx\sqrt{\eta/\xi}d\xi/2$ a generic expression for the ionization rate $W$ in 2D dimensions
is obtained 
\begin{equation}
W=-\frac{i}{\mu}\eta\left(T^{*}\frac{dT}{d\eta}-T\frac{dT^{*}}{d\eta}\right)\int_{0}^{\infty}\frac{\vert v_1(\xi)\vert^{2}}{\xi^{1/2}}d\xi.
\end{equation}
Note the extra factor of $2$ picked up by the symmetric integral in Eq. (\ref{eq:Ionization rate}). Having obtained this expression, we turn our attention to the computation
of $T$.

\subsection{Solution of the tunneling problem using a semiclassical method}

To determine $T(t)$ we need to solve Eq. (\ref{eq:T(t) diff eq}),
which is so far fully equivalent to Eq. (\ref{eq:v1 eta diff eq}).
In order to do so, we will use a uniform JWKB-type solution (where JWKB stands for Jeffrey-Wentzel-Kramers-Brillouin), known
as the Miller and Good approach \citep{miller1953wkb,berry1973semiclassical}.
This method consists of introducing an auxiliary problem, whose solution
is already established, to solve the main equation. The desired wave
function will be given by the product of the solution to the auxiliary problem and a coordinate dependent amplitude. This amplitude consists of the quotient of two functions: in the denominator we have all the
elements of the main equation associated with the non-differentiated
term; in the numerator we have the analogous elements but for the auxiliary
equation. This last term is the key difference between the Miller
and Good approach and the usual JWKB method. While the latter leads to
wave functions with divergences at the classical turning points, the
former produces smooth wave functions across the whole domain. As
the auxiliary problem that will help us solve Eq. (\ref{eq:T(t) diff eq}),
we introduce the Airy equation 
\begin{equation}
\frac{d^{2}}{d\zeta^{2}}\phi(\zeta)+\zeta\phi(\zeta)=0,
\end{equation}
whose solution reads
\begin{equation}
\phi(\zeta)=b_{2}\textrm{Ai}(-\zeta)+b_{1}\textrm{Bi}(-\zeta),\label{eq:Airy solution}
\end{equation}
with $\textrm{Ai}(x)$ and ${\rm Bi}(x)$ the Airy functions. This
equation has a single turning point at $\zeta=0$; the allowed and
forbidden regions are located at $\zeta>0$ and $\zeta<0$, respectively.
In order to have an outgoing wave in the propagating region we need
to choose the coefficients $b_{1}$ and $b_{2}$ in a way that allows
us to recover the correct asymptotic behavior. The asymptotic form
of Eq. (\ref{eq:Airy solution}) reads
\begin{equation}
\phi(\zeta)\xrightarrow[\zeta\rightarrow\infty]{}b_{2}\frac{\sin\left[\frac{\pi}{4}+\frac{2}{3}\zeta^{3/2}\right]}{\sqrt{\pi}\zeta^{1/4}}+b_{1}\frac{\cos\left[\frac{\pi}{4}+\frac{2}{3}\zeta^{3/2}\right]}{\sqrt{\pi}\zeta^{1/4}}.
\end{equation}
In order to obtain a traveling wave we choose $b_{2}=ib_{1}$; with
this choice we obtain 
\begin{equation}
\phi_{\textrm{out}}(\zeta)\sim b_{1}\frac{e^{i\frac{\pi}{4}+i\frac{2}{3}\zeta^{3/2}}}{\sqrt{\pi}\zeta^{1/4}},
\end{equation}
as we desired. When $\zeta\rightarrow-\infty$ the solution Bi($-\zeta$)
grows while $\textrm{Ai}(-\zeta)$ vanishes; thus, deep inside the forbidden
region, we choose to approximate Eq. (\ref{eq:Airy solution}) by
\begin{equation}
\phi_{\textrm{forbidden}}(\zeta)\sim b_{1}\frac{e^{\frac{2}{3}(-\zeta)^{3/2}}}{\sqrt{\pi}(-\zeta)^{1/4}}.
\end{equation}
Using these results we write the solution of (\ref{eq:T(t) diff eq})
in the allowed region as
\begin{equation}
T\left[t(\eta)\right]=b_{1}\left(\frac{\zeta}{P^{2}(t)}\right)^{1/4}\frac{e^{i\frac{\pi}{4}+i\frac{2}{3}\zeta^{3/2}}}{\sqrt{\pi}\zeta^{1/4}},\label{eq:MG allowed region}
\end{equation}
with $P^{2}(t)$ given by Eq. (\ref{eq:P squared def}) and $\zeta(t)$
is defined via the relation
\begin{equation}
\int_{0}^{\zeta}\sqrt{\zeta'}d\zeta'=\frac{2}{3}\zeta^{3/2}=\int_{t_{0}}^{t}\sqrt{P^{2}(t^{\prime})}dt',\label{eq:allowed region zeta def}
\end{equation}
where $t_{0}$ corresponds to the classical turning point of $P^{2}(t)$,
that is $P^{2}(t_{0})=0.$ Combining Eq. (\ref{eq:allowed region zeta def})
with Eq. (\ref{eq:MG allowed region}) it is easily shown that
\begin{equation}
T^{*}\frac{dT}{d\eta}-T\frac{dT^{*}}{d\eta}=\frac{2i}{\pi}\frac{|b_{1}|^{2}}{\eta},
\end{equation}
which produces an ionization rate given by
\begin{equation}
W=\frac{|b_{1}|^{2}}{\pi\mu}\int_{0}^{\infty}\frac{\vert v_1(\xi)\vert^{2}}{\xi^{1/2}}d\xi.
\end{equation}
Thus, to obtain $W$ two tasks remain: find $b_{1}$ and compute the
integral in $\xi$. Let us now focus on the first one and only turn
our attention to the second one later in the text.

\subsection{Matching the wave function to an asymptotic one due to a Coulomb
tail}

In order to obtain $b_{1}$ we follow a conceptually simple procedure:
we will determine the wave function $T\left[t(\eta)\right]$ deep
inside the forbidden region, $-\infty\ll t\ll t_{0}$, in the limit of
a small field $F$, and using the Miller and Good approach we will
extract $b_{1}$ from the comparison of this equation with the asymptotic
solution of the radial Wannier equation.

Once more, using the Miller and Good approach, but this time for the
forbidden region, we write $T\left[t(\eta)\right]$ as

\begin{equation}
T(t)\sim\frac{b_{1}}{\sqrt{\pi}}\left[\frac{1}{-P^{2}(t)}\right]^{1/4}e^{\frac{2}{3}(-\zeta)^{3/2}}.\label{eq:MG forbidden region}
\end{equation}
Note the sign differences between this equation and Eq. (\ref{eq:MG allowed region});
these appear due to the different validity regions of the respective
functions. In the limit of a weak field, $F\rightarrow0$, we can
approximate the $P^{2}(t)$ present in the denominator of the pre-factor
with 
\begin{equation}
-P^{2}(t)\approx-\frac{\mu E}{2}e^{2t}.\label{eq:P^2 approximation}
\end{equation}
The function $\zeta$ is defined by the relation
\begin{equation}
\frac{2}{3}(-\zeta)^{3/2}=\int_{t}^{t_{0}}\sqrt{-P^{2}(t)}dt,\label{eq:zeta forbidden def}
\end{equation}
where once again $t_{0}$ denotes the zero of $P^{2}(t)$, and is
approximately\textbf{ 
\begin{equation}
t_{0}\sim\log\left(-\frac{2E}{F}\right).
\end{equation}
}

To compute the integral in Eq. (\ref{eq:zeta forbidden def}) we recall
the form of $P^{2}(t)$ presented in Eq. (\ref{eq:P squared def})
and discard the term $1/16$, since its contribution
to the overall integral is insignificant. Then, we expand the integrand for small $\lambda,$
a book-keeping multiplicative parameter associated with the potential
and the separation constant. Afterwards, we compute the integral,
return to the original $\eta$ coordinate using the Langer transformation,
and expand the result for small $F$ and large $\eta$, in this order.
\textcolor{black}{We note that the approximation presented in Eq.
(\ref{eq:P^2 approximation}), although suitable for the pre-factor,
is too crude to produce a reasonable result inside the integral.}
Now, the key step in this procedure, is the judicious choice of the
separation constant $Z$; this is a degree of freedom we can take
advantage of by choosing it as we desire. Accordingly, we choose $Z$ as 
\begin{equation}
Z=\frac{\sqrt{-E\mu}}{2\sqrt{2}}.
\end{equation}
This choice is made in order to allow us to recover a function with
an $\eta$ dependence that matches the solution of the asymptotic
differential equation for the Coulomb tail of the interaction potential.
The wave function $u(\eta)$ is obtained from
\begin{equation}
u(\eta)=\eta^{1/4}T(t),
\end{equation}
which can be written as
\begin{align}
u(\eta) & =b_{1}\frac{2^{1+\frac{3\sqrt{-E\mu}}{\sqrt{2}E\kappa}}e^{-\frac{2\sqrt{2}E\sqrt{-E\mu}}{3F}}}{\sqrt{\pi}\mu^{1/4}}(-E)^{\frac{\sqrt{-E\mu}}{\sqrt{2}E\kappa}}F^{-\frac{1}{4}-\frac{\sqrt{-E\mu}}{\sqrt{2}E\kappa}}\label{eq:u_eta_asympt} \notag \\
 & \times e^{-\frac{\eta\sqrt{-E\mu}}{\sqrt{2}}}\eta^{-\frac{1}{2}-\frac{\sqrt{-E\mu}}{\sqrt{2}E\kappa}}. 
%
\end{align}
As briefly explained above, we now compare this wave function with
the asymptotic solution for a particle (the exciton) of energy
$E$ bound by a potential with a Coulomb tail. The asymptotic wave function
in radial coordinates reads 
\begin{align}
\psi_{{\rm asympt}}(r) & \sim Ae^{-r\sqrt{-2E\mu}}r^{\frac{\sqrt{2}\sqrt{\mu}-\sqrt{-E}\kappa}{2\sqrt{-E}\kappa}},\label{eq:psi_r_asympt}
\end{align}
where $A$ is a constant determined from the normalization of the
full wave function due to a Coulomb potential. Note that the wave
function in parabolic coordinates reads $\psi(\eta,\xi)=u(\eta)v(\xi)$.
In the ground state, and due to the symmetry of the equations defining
both $u(\eta)$ and $v(\xi)$ in the absence of the field, we must
have $u(\eta)=v(\eta)$. In the large $\eta$ limit, $\eta\gg\xi$,
we find from Eq. (\ref{eq:psi_r_asympt}) that $u(\eta)$ must be
of the form
\begin{equation}
u(\eta)\sim\sqrt{A}e^{-\frac{\eta}{2}\sqrt{-2E\mu}}(\eta/2)^{\frac{\sqrt{2}\sqrt{\mu}-\sqrt{-E}\kappa}{2\sqrt{-E}\kappa}}\label{eq:u_eta_asympt_2}
\end{equation}
Comparing Eqs. (\ref{eq:u_eta_asympt}) and (\ref{eq:u_eta_asympt_2})
it follows that \textbf{$b_{1}$ }reads 
\begin{align}
b_{1} & =\sqrt{A}\sqrt{\pi}\mu^{1/4}\frac{e^{\frac{2\sqrt{2}E\sqrt{-E\mu}}{3F}}}{F^{-\frac{1}{4}-\frac{\sqrt{-E\mu}}{\sqrt{2}E\kappa}}}2^{-\frac{1}{2}-2\frac{\sqrt{-2E\mu}}{2E\kappa}}(-E)^{-\frac{\sqrt{-2E\mu}}{2E\kappa}}.
\end{align}
Once \textbf{$b_{1}$} has been determined, the remaining task is
the calculation of the integral in the rate equation. Again, we take
advantage of the Coulomb tail present in the potential binding the
exciton. In this case, the radial wave function in a Coulomb potential,
for a particle with energy $E$, reads 
\begin{equation}
R(r)=Ae^{-\sqrt{-2E\mu}r}U\left(-\frac{\sqrt{2\mu}-\sqrt{-E}\kappa}{2\sqrt{-E}\kappa},1,2\sqrt{-2E\mu}r\right).
\end{equation}
where $U(a,b,z)$ is the hypergeometric $U-$function. We now take
$v(\xi)=A^{-1/2}R(\xi/2)$ and perform the integral. The result can
be written as 
\begin{equation}
\int_{0}^{\infty}\left|R(\xi/2)\right|^{2}\frac{d\xi}{\sqrt{\xi}}\equiv AA_{\xi}.
\end{equation}
In general, it follows that the ionization rate $W$ reads 
\begin{align}
W & =\frac{2}{\mu}A_{\xi}A^{2}\sqrt{\mu}\frac{e^{\frac{4\sqrt{2}E\sqrt{-E\mu}}{3F}}}{F^{-\frac{1}{2}-\frac{2\sqrt{-E\mu}}{\sqrt{2}E\kappa}}}2^{-1-4\frac{\sqrt{-2E\mu}}{2E\kappa}}(-E)^{-\frac{\sqrt{-2E\mu}}{E\kappa}} ,\label{eq:Central0} 
%
\end{align}
and for the particular case of a 2D exciton bound by the Coulomb interaction we
obtain 
\begin{equation}
W=32\sqrt{\frac{2}{\pi}}\frac{\mu^{2}e^{-\frac{16\mu^{2}}{3F\kappa^{3}}}}{\sqrt{F}\kappa^{7/2}},\label{eq:H_W}
\end{equation}
a result identical to that found by Tanaka \emph{et al}. \citep{Tanaka1987}.
In general, for any potential with a Coulomb tail we find 

\begin{equation}
W=g_{0}^{2}W_{0}(F),\label{eq:central}
\end{equation}
where
\begin{equation}
W_{0}(F)=\frac{e^{\frac{4\sqrt{2}E\sqrt{-E\mu}}{3F}}}{F^{-\frac{1}{2}-\frac{2\sqrt{-E\mu}}{\sqrt{2}E\kappa}}},
\end{equation}
and

\begin{equation}
g_{0}^{2}\approx\frac{\kappa2^{\frac{2\sqrt{2}\mu}{\kappa\sqrt{-E\mu}}+\frac{5}{4}}(-E){}^{\frac{\sqrt{2}\mu}{\kappa\sqrt{-E\mu}}+\frac{5}{4}}\Gamma\left(\frac{\sqrt{2}\mu}{\kappa\sqrt{-E\mu}}-\frac{1}{2}\right)}{\pi\sqrt[4]{\mu}\Gamma\left(\frac{\sqrt{2}\mu}{\kappa\sqrt{-E\mu}}\right)}\label{eq:g02}
\end{equation}
where $\Gamma(z)$ is the gamma function. Equation (\ref{eq:central})
together with Eq. (\ref{eq:g02}) are the central results of this
paper. Special limits of this last result can be obtained for carefully
chosen values of $E$, $\kappa$, and $\mu$. In particular, for $E$
given by $E=-2\mu/\kappa^{2}$, corresponding to the ground state
energy of a 2D exciton bound by the Coulomb potential, we recover the
result given by Eq. (\ref{eq:H_W}) for the rate $W$. In the next
section we explore the consequences of Eqs. (\ref{eq:central}) and
(\ref{eq:Central0}).

\section{Results\label{sec:Results}}

Having determined the form of the ionization rate, we can compare
our results with numerical ones obtained via the solution of an eigenvalue
problem for the exciton's motion using the complex scaling method,
which allow us to access complex eigenvalues, with the imaginary part
interpreted as the rate $W$ computed above. Below we give a brief
account of how the numerical calculations are performed.

\subsection{Complex scaling method}

When a system that may be ionized is subjected to an external electric
field, the energy eigenvalue turns complex. The ionization rate of
the system is then described by the imaginary part of the energy as $\Gamma = -2\mathrm{Im}\, E$.
A formally exact method of computing the complex energy is to transform
the original eigenvalue problem into a non-hermitian eigenvalue problem
via the complex scaling technique \cite{Balslev1971,Aguilar1971,Herbst1978,Reed1982}.
Here, one rotates the radial coordinate into the complex plane by
an angle $\phi$ to circumvent the diverging behavior of the resonance
states \cite{Ho1983}. The method is incredibly flexible, and one
may choose to either rotate the entire radial domain as $r\to re^{i\phi}$
or choose to only rotate the coordinate outside a desired radius $R$
as 
\begin{align}
r\to\begin{cases}
r\quad & \mathrm{for}\,\,r<R\\
R+\left(r-R\right)e^{i\phi}\quad & \mathrm{for}\,\,r>R\thinspace.
\end{cases}\label{eq:ECStrans}
\end{align}
The latter, so-called exterior complex scaling (ECS) technique, was
first introduced because the uniform complex scaling (UCS) technique
was not applicable within the Born-Oppenheimer approximation \cite{Simon1979}.
When both methods may be applied to the same potential, which is the
case for all potentials considered here, they yield identical eigenvalues.
UCS is easier to implement, and has been used to obtain ionization
rates of excitons in monolayer MoS$_{2}$ \cite{Haastrup2016} and
WSe$_{2}$ \cite{Massicotte:2018aa} for relatively large fields.
However, the ECS technique \cite{Simon1979,Scrinzi1993} is much more
efficient for the weak fields that are relevant for excitons in 2D
semiconductors \cite{Kamban2019}. Using the contour defined by Eq. (\ref{eq:ECStrans})
in the eigenvalue problem, we obtain states that behave completely
differently in the interior $r<R$ and exterior $r>R$ domains. Furthermore,
there are discontinuities at $R$ that we have to deal with \cite{Scrinzi1993,Scrinzi2010}.
An efficient method of solving these types of problems is to use a
finite element basis to resolve the radial behavior of the states.
To this end, we divide the radial domain into $N$ segments $\left[r_{n-1},r_{n}\right]$.
A set of $p$ functions satisfying 
\begin{align}
 & f_{i}^{\left(n\right)}\left(r_{n-1}\right)=f_{i}^{\left(n\right)}\left(r_{n}\right)=0\thinspace,\nonumber \\
\mathrm{except}\quad & f_{1}^{\left(n\right)}\left(r_{n-1}\right)=f_{p}^{\left(n\right)}\left(r_{n}\right)=1\thinspace,\label{ap:FEMECS:eq:cond}
\end{align}
is then introduced on each segment $n$ in order to make enforcing
continuity across the segment boundaries simple. In practice, we transform
the Legendre polynomials such that they satisfy Eq. (\ref{ap:FEMECS:eq:cond})
\cite{Scrinzi2010}. The wave function may then be written as 
\begin{align}
\psi\left(\boldsymbol{r}\right)=\sum_{m=0}^{M}\sum_{n=1}^{N}\sum_{i=1}^{p}c_{i}^{\left(m,n\right)}f_{i}^{\left(n\right)}\left(r\right)\cos\left(m\theta\right)\thinspace,
\end{align}
where continuity across the segment boundaries is ensured by enforcing
\begin{align}
c_{p}^{\left(m,n-1\right)}=c_{1}^{\left(m,n\right)}\thinspace,\quad n=2,...,N\thinspace,\label{eq:cont}
\end{align}
in the expansion coefficients. As the unperturbed problems considered
here are radially symmetric, an efficient angular basis of cosine
functions may be used to resolve the angular behavior of the states.
Using this expansion, the Wannier equation may be transformed into
a matrix eigenvalue problem and solved efficiently using techniques
for sparse matrices. Note that we keep the radial coordinate in the
basis functions real and leave it to the expansion coefficients
to describe the behavior along the complex contour. This technique
has previously been used to compute ionization rates of excitons in
monolayer semiconductors \cite{Kamban2019} as well as bilayer heterostructures
\cite{Kamban2020}, and we shall use it here to validate the analytical
results.

\subsection{An application}

\begin{figure}
\begin{centering}
\includegraphics[width=8cm]{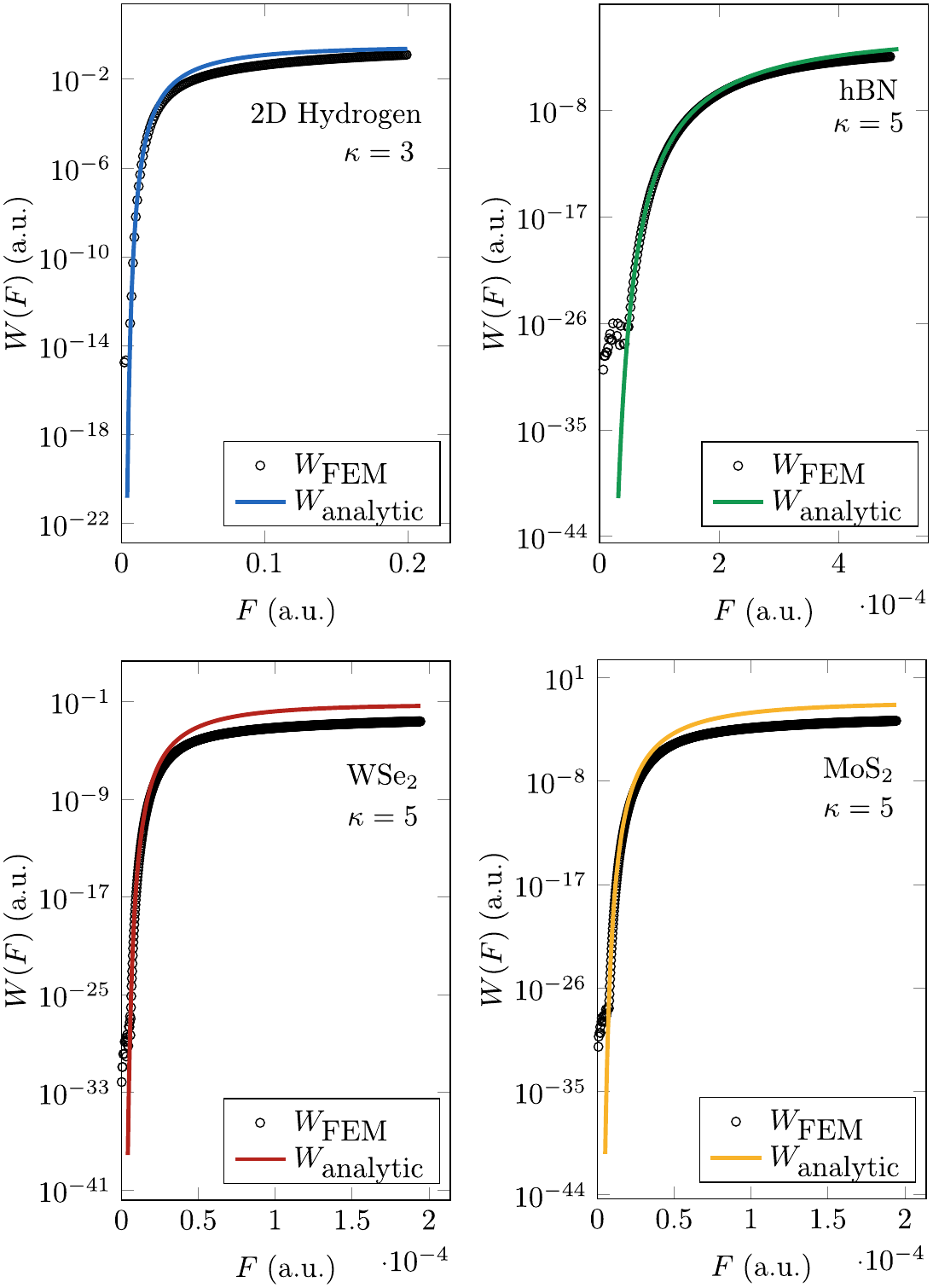} 
\end{centering}
\caption{Comparison of the numerical and analytical calculation of the rate
for hBN excitons. The reduced mass of the exciton in H, hBN, MoS$_2$, WSe$_2$ 
is
$\mu=1,\,0.5,\,0.28,\,0.23$, respectively, the
parameters describing the polarizability of 
hBN, MoS$_2$, WSe$_2$
is 
$r_{0}=10,\,43.4,\,46.2$ ${\rm \text{\AA}}$, respectively,
and the dielectric constants is $\kappa=5$ for all but Hydrogen, for which $\kappa=3$.  All quantities depicted are given in atomic units (a.
u.) and the parameters were taken from Ref. \citep{Kamban2019}.\label{fig:Comparison-of-the-W}}
\end{figure}

To illustrate the validity of our analytical formula over a significant range of values of the external field $F$, we compute the ionization rate for excitons in the 2D Hydrogen atom, hBN, WSe$_2$ and MoS$_2$. In previous
publications we have shown that excitons in hBN and TMDs are well described by
the Wannier equation with the Rytova-Keldysh potential \citep{Ferreira:19,Henriques2020hBN}.
In Fig. \ref{fig:Comparison-of-the-W} we present a comparison of
our analytical results with the finite element method (FEM) approach described above. There
is a remarkable agreement between both approaches across the four cases of study. The analytical
results excel at moderate and small field values, but start to deviate
from the exact numerical methods at extremely large fields. This is to be expected,
since our analytical result was obtained in the limit of small fields.
At very small fields the FEM struggles to give accurate
results, a region where the analytical approach is highly accurate.
Moreover, the FEM requires time-expensive calculations and convergence
needs to be confirmed for every case. Needless to say, the analytical
approach suffers not from these two shortcomings. Also, the analytical
approach makes studying the dependence on the dielectric environment
surrounding the 2D material \citep{Chaves_2017} easy.

\section{Final remarks\label{sec:Conclusions}}

In this paper we have derived an expression for the ionization rate
of excitons in a 2D material due to the application of an external
static electric field. Our result is quantitatively accurate, as was
shown in the bulk of the text. Our approach took a semi-classical
path, based on an approximate separation of the Rytova-Keldysh potential
in parabolic coordinates. This step is key in the derivation, and
is justifiable on the basis of the behavior of the potential near
the origin and at large distances. The next key step is the solution
of a tunneling problem, described by one of the equations arising
from the separability procedure, the $\eta-$equation. The solution
of the tunneling problem was achieved via a uniform semi-classical
method, developed by Miller and Good and used by Berry and Ozorio
de Almeida for the 2D Coulomb problem, for zero angular momentum channel.
Once the semi-classical solution is found, we match it with the asymptotic
solution of a particle of reduced mass $\mu$ (the exciton reduced
mass) and energy $E$, in a dielectric environment characterized by
a dielectric function $\kappa$, in a Coulomb potential. This matching
requires that the original potential binding the electron and hole
has a Coulomb tail, which is fortunately true in our case. Therefore,
for every potential with a Coulomb tail our method is applicable.
One advantage of the method, besides giving an analytical solution
for the ionization rate, is that it is easily extendable to other
classes of potentials such as those discussed by Pfeiffer \citep{Pfeiffer2012}.
Finally, we note that our result can be extended to the calculation
of the photo-ionization rate of the exciton due to an external electric
field of frequency $\omega$. To achieve this, we replace the electric
field strength by $F(t)=F_{0}\cos(\omega t)$ in the rate equation
and average over one cycle. Although this procedure is not exact,
it should give good results in the low frequency regime. 
\begin{acknowledgments}
N.M.R.P. acknowledges support from the European Commission through
the project ``Graphene-Driven Revolutions in ICT and Beyond'' (Ref.
No. 881603 -{}- core 3), and the Portuguese Foundation for Science
and Technology (FCT) in the framework of the Strategic Financing UID/FIS/04650/2019.
In addition, N. M. R. P. acknowledges COMPETE2020, PORTUGAL2020, FEDER
and the Portuguese Foundation for Science and Technology (FCT) through
projects POCI- 01-0145-FEDER-028114, POCI-01-0145-FEDER- 029265, PTDC/NAN-OPT/29265/2017,
and POCI-01-0145-FEDER-02888. 
H.C.K and T.G.P gratefully acknowledge financial support from the Center for Nanostructured Graphene (CNG), which is sponsored by the Danish National Research Foundation, Project No. DNRF103. Additionally, T.G.P. is supported by the QUSCOPE Center, sponsored by the Villum Foundation.
\end{acknowledgments}


%

\end{document}